\documentclass[aps,PRResearch,twocolumn,longbibliography,showpacs,showkeys,footinbib,groupedaddress]{revtex4-2}

\usepackage{amsmath}
\usepackage{graphicx}
\usepackage{bm}
\usepackage{color}
\usepackage{epstopdf}

\usepackage{hyperref}
\usepackage[all]{hypcap}

\renewcommand{\i}{\ensuremath{\mathrm{i}}}
\newcommand{\e}{\ensuremath{\mathrm{e}}}

\begin{document}

\title{Crystalline Anisotropic Topological Superconductivity in Planar Josephson Junctions}

\author{
Joseph D. Pakizer$^{1}$,~Benedikt Scharf$^{2}$,~Alex Matos-Abiague$^{1}$
}

\affiliation{
$^{1}$\textit{Department of Physics \& Astronomy, Wayne State University, Detroit, MI 48201, USA}\\
$^{2}$\textit{Institute for Theoretical Physics and Astrophysics, University of W\"{u}rzburg, Am Hubland, 97074 W\"{u}rzburg, Germany}
}

\date{\today}
 
\begin{abstract}
We theoretically investigate the crystalline anisotropy of topological phase transitions in phase-controlled planar Josephson junctions (JJs) subject to spin-orbit coupling and in-plane magnetic fields. It is shown how topological superconductivity (TS) is affected by the interplay between the magnetic field and the orientation of the junction with respect to its crystallographic axes. This interplay can be used to electrically tune between different symmetry classes in a controlled fashion and thereby optimize the stability and localization of Majorana bound states in planar Josephson junctions. Our findings can be used as a guide for achieving the most favorable conditions when engineering TS in planar JJs and can be particularly relevant for setups containing non-collinear junctions which have been proposed for performing braiding operations on multiple Majorana pairs.
\end{abstract}

\maketitle

\section{Introduction}

Majorana bound states (MBS) are localized zero-energy quasiparticle excitations at the boundaries of topological superconductors~\cite{Kitaev2001:PU,Alicea2012:RPP,Leijnse2012:SST,Beenakker2013:ARCMP,Aguado2017:RNC}. These states are not only of tremendous interest for fundamental research, but also because their non-Abelian statistics makes them ideal building blocks for fault tolerant quantum computation~\cite{Nayak2008:RMP,Kitaev2003:AP,Alicea2011:NP}. Realizations~\cite{Mourik2012:S,Das2012:NP,Deng2012:NL,Deng2016:S,NadjPerge2014:S} of topological superconductors hosting MBS are usually sought in materials with proximity-induced $s$-wave pairing and a nontrivial spin structure, typically provided by spin-orbit coupling (SOC) and/or magnetic textures~\cite{Fu2008:PRL,Lutchyn2010:PRL,Oreg2010:PRL,Pientka2012:PRL,Klinovaja2012:PRL,Dominguez2017:NPJQM,Fleckenstein2018:PRB,Kjaergaard2012:PRB,Mohanta2019:PRA,Schuray2020:EPJST}. In the pursuit of topological superconductivity (TS), early experimental efforts have focused mostly on one-dimensional (1D) systems such as hybrid structures of superconductors and semiconductor nanowires~\cite{Mourik2012:S,Das2012:NP,Deng2012:NL,Deng2016:S} or atomic chains~\cite{NadjPerge2014:S}. Although there is mounting evidence pointing to the appearance of MBS in such 1D systems, a major challenge in the field is to find flexible alternative platforms that do not require fine-tuning of parameters, can be easily scaled to large numbers of states, and enable the implementation of braiding protocols.

\begin{figure}[t]
\centering
\includegraphics*[width=8.5cm]{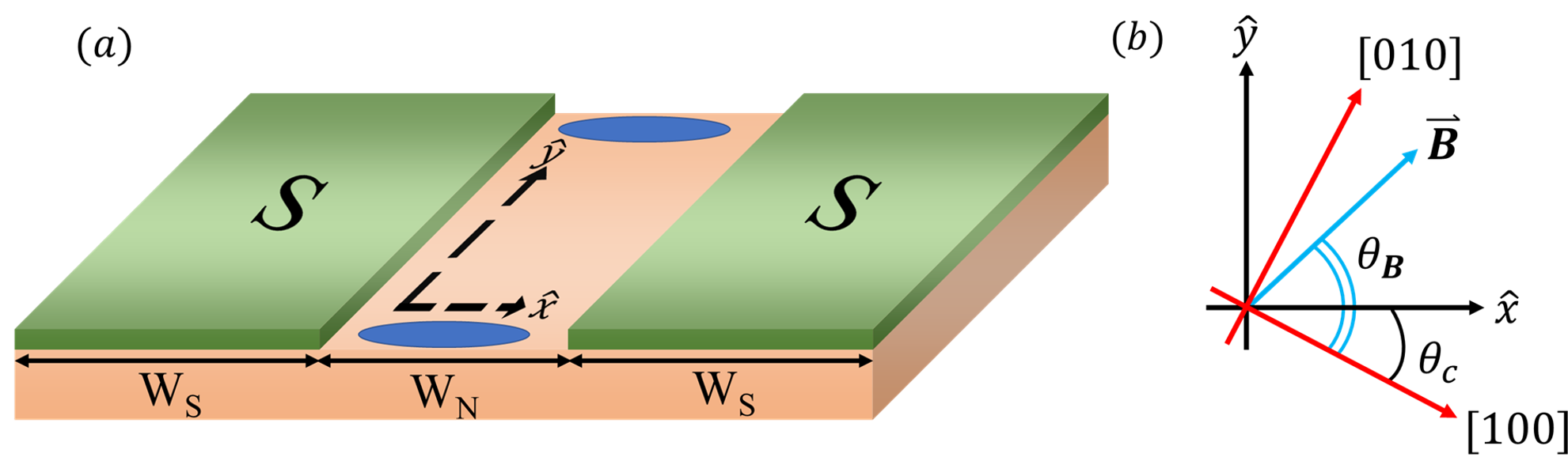}
\caption{(a) A JJ composed of a non-centrosymmetric semiconductor 2DEG in contact to two superconducting (S) leads. The $\hat{x}$ and $\hat{y}$ axes define the coordinate system in the junction reference frame. The blue areas represent MBS. The Rashba SOC strength can be controlled by using a gate on the top of the normal region~\cite{Mayer2021:PRL,Mayer2020:NC}. (b) $\theta_B$ and $\theta_c$ characterize the orientation of the in-plane magnetic field ($\mathbf{B}$) and the junction reference frame, respectively, with respect to the semiconductor [100] crystallographic axis.
}\label{fig1:syst}
\end{figure}

A promising route to address these issues is to go to two-dimensional (2D) geometries, especially in light of the remarkable experimental progress in proximity-inducing superconductivity in 2D systems and surface states~\cite{Hart2014:NP,Wan2015:NC,Shabani2016:PRB,Kjaergaard2016:NC,Suominen2017:PRL,Hart2017:NP,Maier2012:PRL,Sochnikov2015:PRL}. Among the various proposals for 2D setups hosting MBS~\cite{Fu2008:PRL,Sau2010:PRL,Alicea2010:PRB,Fatin2016:PRL,MatosAbiague2017:SSC,Zhou2019:PRB,Pientka2017:PRX,Hell2017:PRL,Melo2019:SP}, those based on phase-controlled planar Josephson junctions (JJs) [Fig.~\ref{fig1:syst}(a)] appear particularly auspicious~\cite{Fu2008:PRL,Pientka2017:PRX,Hell2017:PRL}. In fact, there is already tentative evidence for a topological phase transition and TS in such semiconductor-based JJs~\cite{Fornieri2019:N,Ren2019:N,Mayer2021:PRL}. There has, however, been no conclusive experimental evidence of MBS in planar JJs yet. Hence, finding conditions under which well-localized MBS form in planar JJs is a topic studied vigorously~\cite{Virtanen2018:PRB,Setiawan2019:PRB2,Laeven2020:PRL,Liu2019:PRB,Haim2019:PRL,Setiawan2019:PRB,Scharf2019:PRB}.

Most theoretical works~\cite{Pientka2017:PRX,Hell2017:PRL,Setiawan2019:PRB2,Laeven2020:PRL,Liu2019:PRB,Haim2019:PRL,Setiawan2019:PRB,Stern2019:PRL,Zhou2019:PRL} on planar JJs have considered the effects of Rashba SOC resulting from structure inversion asymmetry~\cite{Zutic2004:RMP,Fabian2007:APS} but have ignored Dresselhaus SOC intrinsically present in non-centrosymmetric semiconductors due to the lack of bulk inversion symmetry~\cite{Zutic2004:RMP,Fabian2007:APS}. Without Dresselhaus SOC, the Rashba SOC field exhibits a $C_\infty$ (or $C_4$ if contributions cubic in momentum are considered) symmetry. However, the presence of both Dresselhaus and Rashba SOCs lowers the symmetry to $C_{2v}$, resulting in various magnetoanisotropic phenomena in both the normal~\cite{Moser2007:PRL,Badalyan2009:PRB,Matos-Abiague2015:PRL} and superconducting~\cite{Ikegaya2017:PRB,Biderang2018:PRB,Hogl2015:PRL,Costa2019:PRB,Alidoust2020:PRB} states. Magnetoanisotropic effects due to the co-existence of Rashba and Dresselhaus SOC in planar JJs~\cite{Scharf2019:PRB,Alidoust2021:arXiv} and their relevance for the realization of TS have recently been theoretically investigated~\cite{Scharf2019:PRB}.

In addition to magnetoanisotropy, crystalline anisotropic effects have also been observed in systems with coexisting Rashba and Dresselhaus SOCs in the normal state~\cite{Hupfauer2015:NC}. Here we theoretically investigate crystalline anisotropic TS (CATS) in a planar JJ, i.e., how TS is affected by the orientation of the junction with respect to a fixed crystallographic axis. The realization of TS strongly depends on both the crystallographic orientation of the junction and the direction of the applied magnetic field. Therefore, understanding the properties of CATS is crucial for the optimal experimental design of planar JJs. Furthermore, in dependence of the crystallographic orientation, a top gate tuning the Rashba SOC strength can be used for controlling TS \cite{Mayer2021:PRL,Mayer2020:NC}.

\section{Theoretical Model}
We consider a planar JJ composed of a 2D electron gas (2DEG) formed in a non-centrosymmetric semiconductor and subject to an in-plane magnetic field $\mathbf{B}$ (Fig.~\ref{fig1:syst}). Superconducting regions (S) are induced in the 2DEG by proximity to a superconducting cover, such as Al or Nb, while the uncovered region remains in the normal (N) state. The system is described by the Bogoliubov-de Gennes (BdG) Hamiltonian
\begin{equation}\label{H-BdG}
H=H_{0}\tau_z-\frac{g^\ast\mu_B}{2}\mathbf{B}\cdot\boldsymbol{\Sigma}+\Delta(x)\tau_+ +\Delta^\ast(x)\tau_-\;,
\end{equation}
where
\begin{eqnarray}\label{Ho}
H_{0}&=&\frac{\mathbf{p}^2}{2m^\ast}+V(x)-(\mu_S-\varepsilon)+\frac{\alpha}{\hbar}\left(p_y\sigma_x - p_x\sigma_y\right)\\ \nonumber
&+&\frac{\beta}{\hbar}[(p_x\sigma_x - p_y\sigma_y)\cos2\theta_c - (p_x\sigma_y + p_y\sigma_x)\sin2\theta_c]\;,
\end{eqnarray}
and $\sigma_{x,y,z}$ and $\tau_{x,y,z}$ represent Pauli matrices in spin and Nambu space respectively with $\tau_\pm =(\tau_x\pm\tau_y)/2$. Here $\mathbf{p}$ is the momentum, $m^\ast$ the electron effective mass, $\alpha$ and $\beta$ are, respectively, the Rashba and Dresselhaus SOC strengths, $\theta_c$ characterizes the junction orientation with respect to the [100] crystallographic direction of the semiconductor [Fig.~\ref{fig1:syst}(b)], and $V(x)=(\mu_S - \mu_{N})\Theta(W_N/2-|x|)$ describes the difference between the chemical potentials in the N ($\mu_N$) and S ($\mu_S$) regions. The chemical potentials are measured with respect to the minimum of the single-particle energies, $\varepsilon=m^\ast\lambda^2(1+\left|\sin 2\theta_c\right|)/2\hbar^2$, where we have used the SOC parametrization
\begin{equation}\label{param}
\alpha=\lambda\cos\theta_{so}\;,\;\;\beta=\lambda\sin\theta_{so}\;,\;\;\lambda=\sqrt{\alpha^2+\beta^2}.
\end{equation}

The second contribution in Eq.~(\ref{H-BdG}), with the Dirac spin matrices $\boldsymbol{\Sigma}=\boldsymbol{\sigma}\tau_0$, represents the Zeeman splitting due to an applied magnetic field,
\begin{equation}\label{def-B}
\mathbf{B}=|\mathbf{B}|
\begin{pmatrix}
\cos(\theta_B - \theta_c) \\
\sin(\theta_B - \theta_c) \\
0
\end{pmatrix},
\end{equation}
whose direction with respect to the [100] crystallographic direction is given by $\theta_B$ [Fig.~\ref{fig1:syst}(b)]. The spatial dependence of the superconducting gap is $\Delta(x)=\Delta \e^{i\,{\rm sgn}(x)\phi/2}\Theta(|x|-W_N/2)$, where $\phi$ is the phase difference accross the JJ.

\begin{table}
	\caption{Parameter space for which $H$ belongs to the BDI symmetry class. $n$ represents an integer number.} \label{tab}
	\begin{ruledtabular}
		\begin{tabular}{ccccc}
			$\alpha$ & $\beta$ & $\theta_c$ & $\theta_B$ & $\varphi$\\
			\hline
			$\neq 0$ & 0 & any & $\theta_c +\frac{(2n+1)\pi}{2}$ & $n\pi$ \\
			\hline
			0 & $\neq 0$ & any & $n\pi-\theta_c$ & $\frac{(2n+1)\pi}{2} - 2\theta_c$ \\
			\hline
			$\neq 0$ & $\neq 0$ & $\frac{(2n+1)\pi}{4}$ & $\theta_c +\frac{(2n+1)\pi}{2}$ & $n\pi$ \\
		\end{tabular}
	\end{ruledtabular}
\end{table} 

\section{Symmetry Analysis}

The BdG Hamiltonian~(\ref{H-BdG}) anticommutes with the charge conjugation operator $\mathcal{C}=\sigma_y\tau_y\mathcal{K}$ ($\mathcal{C}^2=1$), as a manifestation of the particle-hole symmetry. The presence of $\mathbf{B}$ and/or $\phi$ breaks the conventional time-reversal symmetry and $[H,T]\neq 0$, where $T=-\i\sigma_y\mathcal{K}$, and $\mathcal{K}$ indicates complex conjugation. Therefore, Eq.~(\ref{H-BdG}) belongs, generically, to symmetry class D. However, under some conditions a transition to the higher BDI symmetry class can occur when an effective time-reversal symmetry emerges in the system. In the absence of Dresselhaus SOC, the symmetry properties of the system are mainly determined by the relative orientation of the magnetic field with respect to the junction direction. In such a case the topological superconducting state becomes magnetoanisotropic but crystalline anisotropy is absent. Assuming a symmetric junction (i.e., if the superconducting leads are identical) oriented along the $y$ axis [see Fig.1(a)], it is possible to define an \emph{effective} time-reversal operator \cite{Pientka2017:PRX},
\begin{equation}\label{t-eff-r}
\mathcal{T}=i\;\Sigma_x \mathcal{R}_x T
\end{equation}
that performs a conventional time-reversal operation, followed by space and spin reflections [$\mathcal{R}_x=(x\rightarrow -x)$ and $i\;\Sigma_x$, respectively] with respect to the $yz$ plane. Note that like the conventional time-reversal operator, $T$, the operator in Eq.~(\ref{t-eff-r}) is anti-unitary and satisfies $\mathcal{T}^2=1$. For this reason we refer to it as an \emph{effective} time-reversal operator.

The spin-orbit field resulting from the combination of Rashba and Dresselhaus SOCs is no longer rotational invariant. Therefore, in order to investigate the effects of crystalline anisotropy on the symmetry properties of the system we need to generalize the definition of the effective time-reversal operator given in Eq.~(\ref{t-eff-r}). Since we choose the coordinate system in such a way that the $x$ axis is always perpendicular to the junction direction (see Fig.1), we can keep the presence of the space reflection $\mathcal{R}_x$ in the generalized definition of $\mathcal{T}$. However, the Rashba+Dresselhaus spin-orbit field strongly depends on the crystallographic orientation of the junction [the dependence of the SOC on $\theta_c$ is quite apparent in Eq.~(\ref{Ho})], hence the spin reflection with respect to the $yz$ plane must be replaced by a spin reflection that can account for different reflection planes containing the crystallographic direction of the junction. Those spin reflection planes can be obtained by rotation of the $yz$ plane about the $z$ axis and are therefore described by the operator,
\begin{equation}
i(\mathbf{n}\cdot\boldsymbol{\Sigma})=i(\cos\varphi\;\Sigma_x+\sin\varphi\;\Sigma_y),
\end{equation}
where $\varphi$ is an angle parameterizing the direction of the unit vector normal to the plane, $\mathbf{n}=(\cos\varphi,\sin\varphi,0)^T$. Therefore, in the most general case, we can define the effective time-reversal operator as,
\begin{equation}\label{tr-op}
\mathcal{T}=i\left(\mathbf{n}\cdot\boldsymbol{\Sigma}\right)\mathcal{R}_x T=i(\cos\varphi\Sigma_x+\sin\varphi\Sigma_y)\mathcal{R}_x T\;.
\end{equation}
Since $\mathcal{T}^2=1$, by requiring $[H,\mathcal{T}]=0$ one can determine the regions of the ($\alpha,\beta,\theta_c,\theta_B$) parameter space for which Eq.~(\ref{H-BdG}) belongs to the BDI symmetry class, independently of $\phi$. When the Hamiltonian belongs to the BDI symmetry class, it also possesses chiral symmetry, $\{\mathcal{S},H\}=0$, characterized by the chiral operator $\mathcal{S}=\mathcal{C}\mathcal{T}$ ($\mathcal{S}^2=1$). The results of the symmetry analysis are shown in Table~\ref{tab}. The presence of SOC leads to the magnetoanisotropy of the topological state and the BDI class emerges only for specific directions of $\mathbf{B}$ with respect to the junction orientation. Furthermore, when only one type of SOC is present, the BDI symmetry class can always be achieved (as long as $\mathbf{B}$ is properly oriented), independently of the junction orientation. For $\alpha\neq 0$, $\beta=0$, and $\theta_c=0$ we recover the results reported in Ref.~\cite{Pientka2017:PRX}. However, the coexistence of Rashba and Dresselhaus SOC results in crystalline anisotropy and reduces the parameter space of the BDI class, which in such circumstances can only occur when $\theta_c$ equals an odd multiple of $\pi/4$, i.e., when the junction orientation is aligned with one of the symmetry axes of the total SOC field pointing along the $[110]$ and $[\bar{1}10]$ crystallographic directions of the proximitized semiconductor. This is a distinctive property of CATS which, as explained below, can be used for removing inconvenient BDI subclasses from the class D phase.

\section{Topological Gap and Topological Charge}
To better understand the magneto-crystalline anisotropy of the TS phase and the symmetry classes, we calculate the topological gap
\begin{equation}\label{topogap}
\Delta_{\rm top}=\min_{k_y}|E(k_y)|,
\end{equation}
for a system with translational invariance along the junction direction (i.e., the $y$ direction). In such a system the momentum component $p_y$ can be substituted by $\hbar k_y$ in Eq.~(\ref{H-BdG}) and we compute its Andreev spectrum $E(k_y)$ numerically for all $k_y$. Then $\Delta_{\rm top}$ is obtained as the eigenenergy closest to zero, as indicated by Eq.~(\ref{topogap}). The size of $\Delta_{\rm top}$ determines the degree of topological protection of the TS state and can be related to the localization of the MBS that would emerge if the system were also confined to finite length in the junction direction.

Complementary to $\Delta_{\rm top}$, we calculate the topological charge $Q$ (i.e., the $Z_2$ topological index associated to symmetry class D),
\begin{equation}\label{qtopo}
Q = {\rm sgn}\left[\frac{{\rm Pf}\{H(k_y = \pi)\sigma_y\tau_y\}}{{\rm Pf}\{H(k_y = 0)\sigma_y\tau_y\}}\right],
\end{equation}
where ${\rm Pf}\{...\}$ denotes the Pfaffian~\cite{Tewari2012:PRL}. $Q$ determines whether the system is in a trivial ($Q=1$) or topological ($Q=-1$) phase.

\begin{figure}[t]
\centering
\includegraphics*[width=8.5cm]{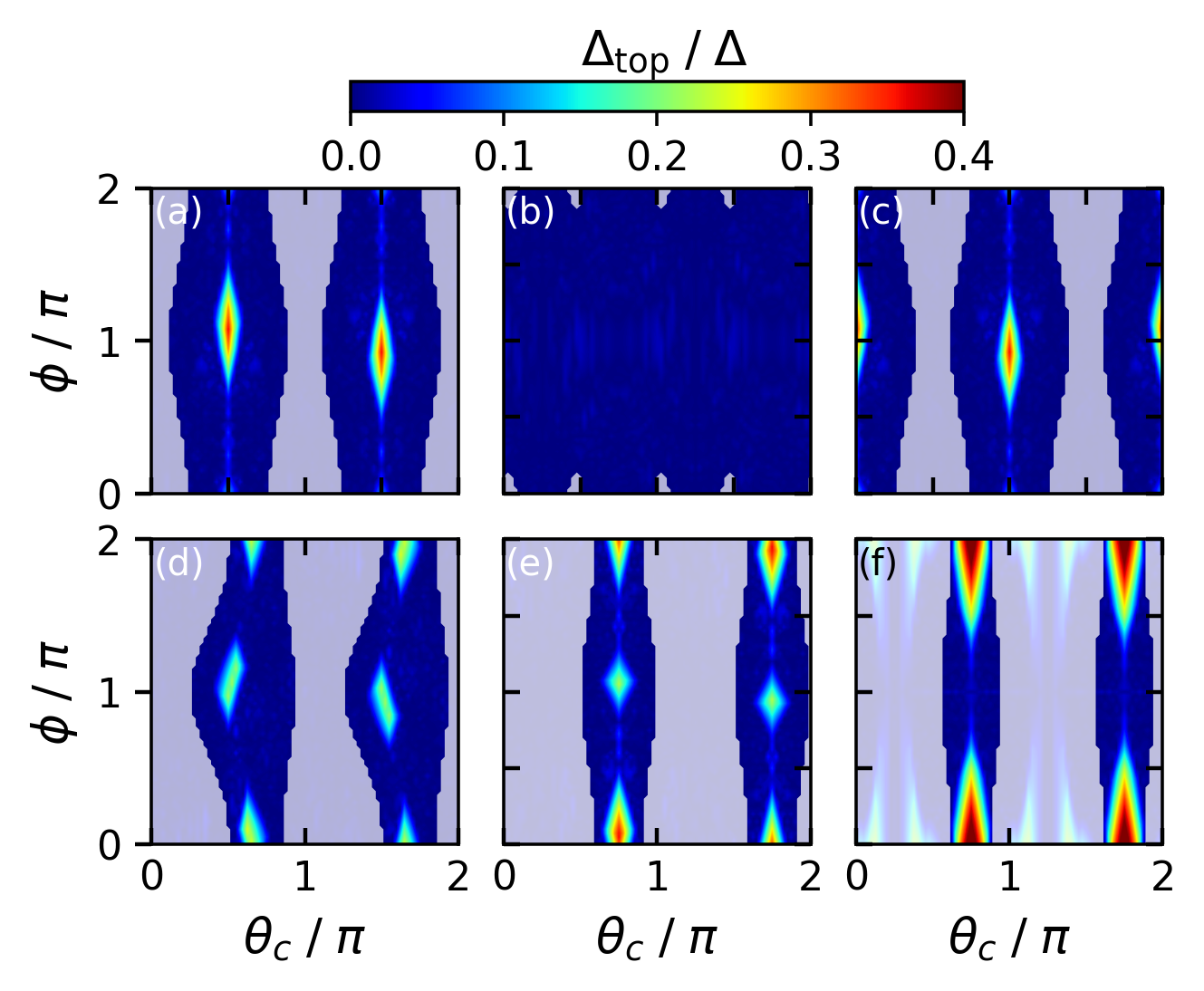}
\caption{Topological gap $\Delta_{\rm top}$ as a function of $\theta_c$ and $\phi$ for (a) $\theta_{so}=0$ (only Rashba SOC) and $\theta_B=0$, (b) $\theta_{so}=\pi/4$ and $\theta_B=0$, (c) $\theta_{so}=\pi/2$ (only Dresselhaus SOC) and $\theta_B=0$, (d) $\theta_{so}=\theta_B=\pi/8$, (e) $\theta_{so}=\pi/8$ and $\theta_B=\pi/4$, and (f) $\theta_{so}=\theta_B=\pi/4$. Gray-shaded and non-shaded areas represent trivial ($Q=1$) and class D TS ($Q=-1$) respectively, except along the thin vertical traces appearing in (a), (c), (e), and (f), which correspond to the BDI topological phase. The BDI class emerges when the conditions in Table~\ref{tab} are fulfilled.}\label{fig:thc-phi}
\end{figure}

\section{Numerical Simulations}
For illustration, we performed numerical simulations using the Kwant package~\cite{Groth2014:NJP} and a discretized version of Eq.~(\ref{H-BdG}) with lattice constant $a=20$~nm. We chose system parameters similar to those found in Al/InAs$_{1-x}$Sb$_x$ JJs~\cite{Mayer2020:AEM}, namely: $m^\ast=0.013~m_0$ (with $m_0$ the bare electron mass), $\Delta = 0.21$~meV, $g^\ast = -20$, and $\lambda=15$~meV~nm. Moreover, $B=0.6$~T, $\mu_S = \mu_N = 2$~meV, $W_S = 350$~nm, and $W_N=100$~nm.

The dependence of $\Delta_{\rm top}$ on $\theta_c$ and $\phi$ is shown in Fig.~\ref{fig:thc-phi} for different $\theta_B$ and different ratios of Rashba vs Dresselhaus, parametrized by the angle $\theta_{so}$ ($\cot\theta_{so}=\alpha/\beta$). Figure~\ref{fig:thc-phi} reveals that it is possible to design JJs, which in dependence on $\theta_c$ and by properly tuning $\theta_{so}$ and $\theta_B$, become topological with a sizable gap at $\phi$ around both $0$ and $\pi$ [Figs.~\ref{fig:thc-phi}(d,e)] or only around $\phi=0$ [Fig.~\ref{fig:thc-phi}(f)] or $\phi=\pi$ [Figs.~\ref{fig:thc-phi}(a,c)].

When the conditions in Table~\ref{tab} are fulfilled, BDI-class subregions appear in the form of vertical traces inside the topological area, at $\theta_c=(2n+1)\pi/2$ and $n\pi$ [Figs.~\ref{fig:thc-phi}(a,c)] and at $\theta_c=(2n+3)\pi/4$ [Figs.~\ref{fig:thc-phi}(e,f)]. For the parameters used in Figs.~\ref{fig:thc-phi}(b,d) the conditions in Table~\ref{tab} are not met and no BDI traces form. Along the BDI-class traces, multiple gap closings and reopenings occur, indicating topological phase transitions between regions with different odd values of the $Z$ invariant~\cite{Pientka2017:PRX}. Therefore, multiple MBS may appear at each end of a confined junction when the system is in the BDI phase. To have a single and stable MBS at each end of the junction, it is convenient to lower the symmetry in a controllable way and leave the system in the D phase~\cite{Pientka2017:PRX,Setiawan2019:PRB2}. This can be done by tuning $\theta_B$ away from the conditions given in Table~\ref{tab}. However, such a detuning leads to a rapid decrease of $\Delta_{\rm top}$. This is apparent in Fig.~\ref{fig:thc-phi}, where the system becomes practically gapless away from the BDI-class traces. However, due to the restrictive character of CATS with respect to the BDI phase, for certain junction and magnetic field orientations, a proper tuning of $\theta_{so}$ can suppress the BDI class and lower the symmetry of the whole topological region to class D, while maintaining a sizable $\Delta_{\rm top}$ for $\phi$ around $0$ or $\pi$ [Fig.~\ref{fig:thc-phi}(d)].  Figure~\ref{fig:thc-phi} illustrates how $\theta_{so}$ and $\theta_{B}$ can be used as tuning knobs for the CATS phase, enhancing system control of topological phase transitions between trivial, class BDI and gapped D phases in planar JJs.


\begin{figure}[t]
\centering
\includegraphics*[width=8.5cm]{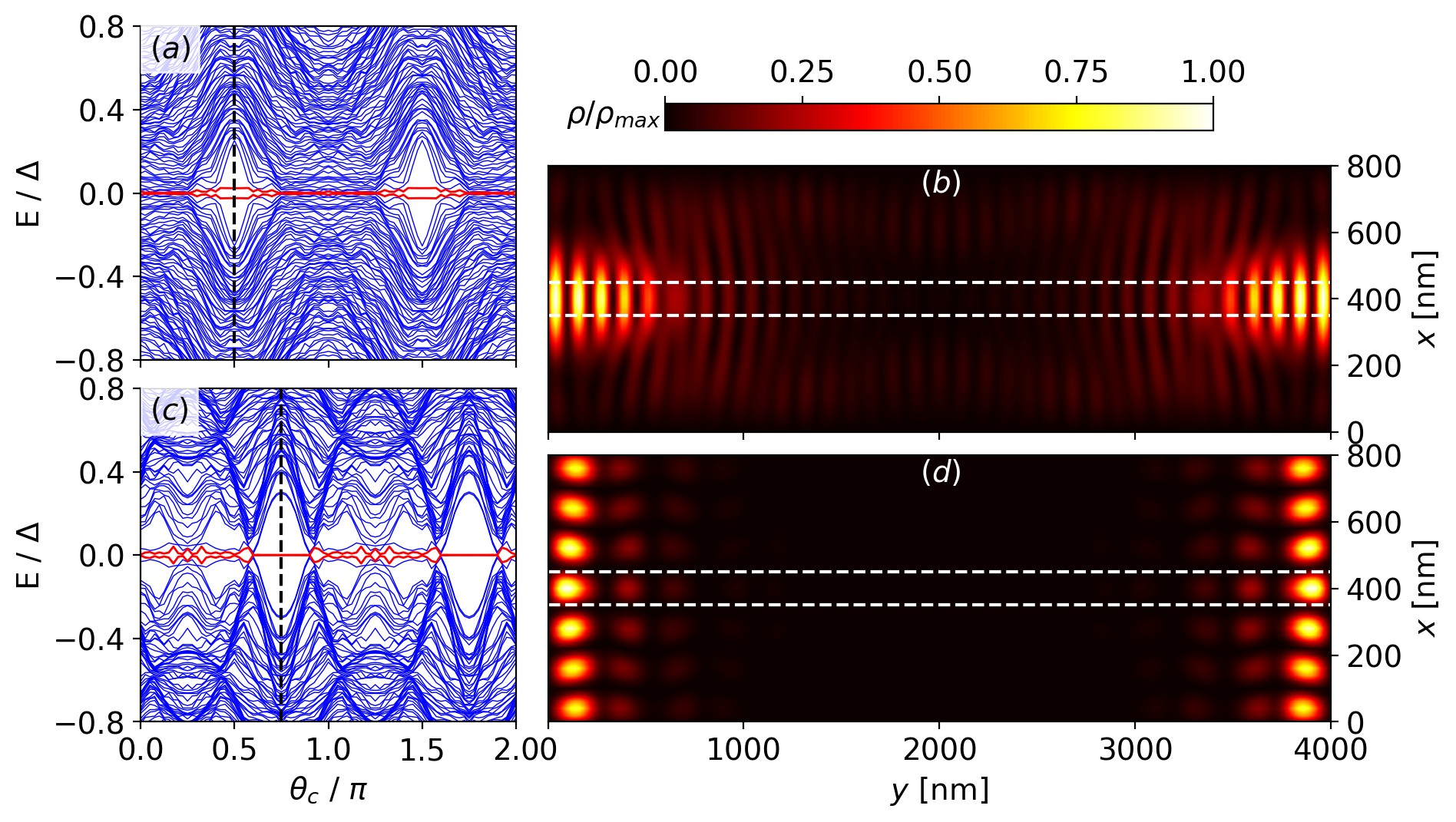}
\caption{Energy spectrum as a function of $\theta_c$ for (a) $\theta_{so}=0$ (only Rashba SOC), $\theta_B=0$, and $\phi=\pi$, and (c) $\theta_{so}=\theta_B=\pi/4$ and $\phi=0$. Red lines indicate the states with energies closest to zero, while black dashed vertical lines indicate the crystallographic orientation resulting in a maximal $\Delta_{\rm top}$. (b) and (d) show the probability density (normalized to its maximum value) of the MBS in (a) and (c), respectively, at $\theta_c$ values specified by the vertical dashed lines.}\label{fig:spectrum}
\end{figure}

\subsection{Energy Spectrum and MBS Localization}
The energy spectrum as a function of $\theta_c$ is shown in Figs.~\ref{fig:spectrum}(a,c) for junctions with Rashba and Rashba+Dresselhaus SOC, respectively. For Fig.~\ref{fig:spectrum}(a) we used the same set of parameters as in Fig.~\ref{fig:thc-phi}(a) and $\phi=\pi$, while for Fig.~\ref{fig:spectrum}(c) the same parameters as in Fig.~\ref{fig:thc-phi}(f) and $\phi=0$ were chosen. Crystalline anisotropic effects on the energy spectrum are apparent in Figs.~\ref{fig:spectrum}(a,c), where the spectra are gapless for most junction orientations, except for values of $\theta_c$ close to satisfying the conditions in Table~\ref{tab}. Note that the behavior of $\Delta_{\rm top}$ shown in Figs.~\ref{fig:spectrum}(a,c) is in perfect agreement with the predictions of Figs.~\ref{fig:thc-phi}(a,f) at $\phi=\pi$ and $\phi=0$, respectively.

MBS probability densities for junction orientations corresponding to red and black intersections in the spectra of Figs.~\ref{fig:spectrum}(a,c) are shown in Figs.~\ref{fig:spectrum}(b,d), respectively. When only Rashba (or only Dresselhaus; see Appendix B) SOC is present, the formation of MBS with finite $\Delta_{\rm top}$ is favored when $\phi=\pi$. However, in the presence of crystalline anisotropy (i.e., $\alpha\neq0$ and $\beta\neq0$), well-localized MBS with a sizable $\Delta_{\rm top}$ are possible at $\phi=\pi$ and $\phi=0$, as shown in Fig.~\ref{fig:spectrum}(d). Interestingly, when $\phi=0$, the MBS localize along the edges perpendicular to the junction [Fig.~\ref{fig:spectrum}(d)], while for $\phi=\pi$ the states are mainly localized at the end regions, inside the junction [Fig.~\ref{fig:spectrum}(c)]. Evidence of edge MBS has also been provided by model calculations considering a narrow junction with Rashba SOC and $\mathbf{B}$ present only in the normal region \cite{Fornieri2019:N}. Although most of previous investigations have focused on the formation of end MBS at $\phi=\pi$, to the best of our knowledge, the existence of edge MBS in narrow junctions where $\mathbf{B}$ extends over the whole system \cite{note_Fraunhofer} has not been previously discussed. Our calculations indicate that such states can naturally emerge using CATS, where a sizable $\Delta_{\rm top}$ can develop at $\phi=0$. A qualitative analysis of the magneto-crystalline anisotropic effects on the localization length of the MBS is given in the Appendix A.

\begin{figure}[t]
\centering
\includegraphics*[width=8.5cm]{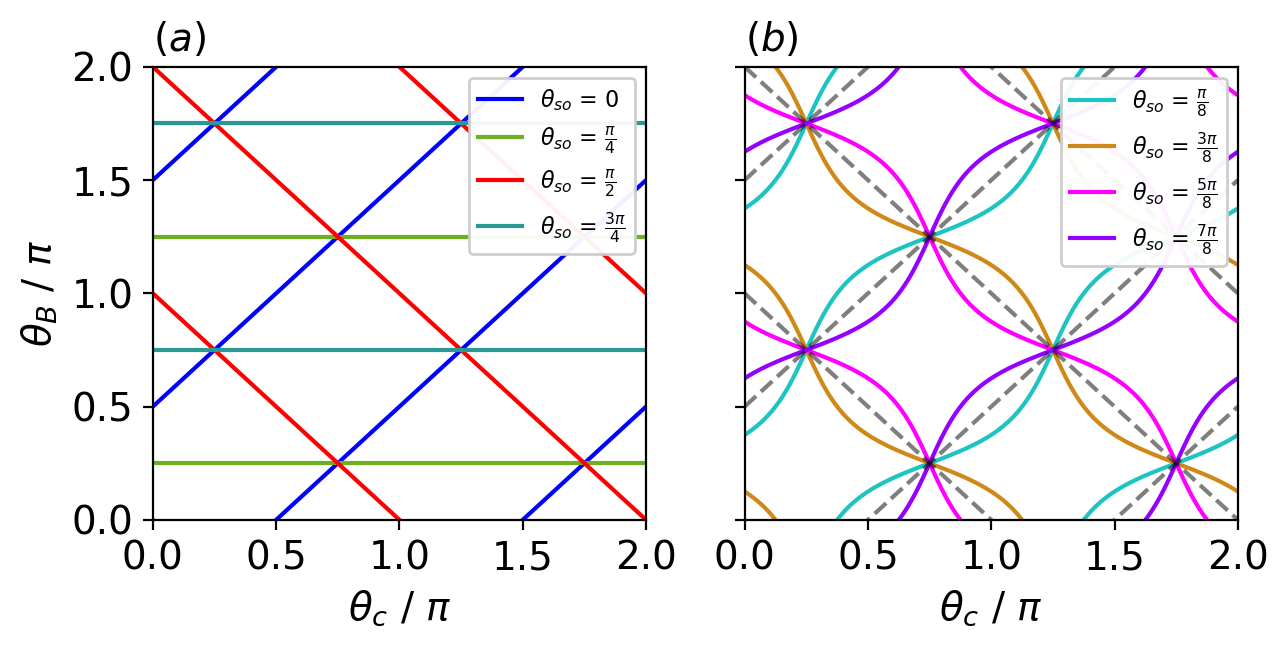}
\caption{Optimal magnetic field direction $\theta_B$ as a function of $\theta_c$. (a) $\theta_{so} = \{0, \frac{\pi}{4}, \frac{\pi}{2}, \frac{3\pi}{4}\}$ (here BDI and optimal conditions coincide). (b) $\theta_{so} = \{\frac{\pi}{8}, \frac{3\pi}{8}, \frac{5\pi}{8}, \frac{7\pi}{8}\}$. Dashed lines with positive (negative) slope correspond to the BDI-class condition when only Rashba (Dresselhaus) SOC is present (Table~\ref{tab}). The intersection points represent the BDI-class condition when both Rashba and Dresselhaus SOCs are present.}\label{fig:optimal}
\end{figure}

\subsection{Optimal Magnetic Field Orientation}

As illustrated in Fig.~\ref{fig:thc-phi}, although the topological region in the parameter space is fairly large, a sizable $\Delta_{\rm top}$ exists only in reduced subregions. Therefore, knowing the conditions leading to a finite  $\Delta_{\rm top}$ as large as possible is crucial for finding well-localized MBS.

In order to find the conditions for reaching the optimal topological gap we consider a junction with translational invariance along the $y$ axis. In such a case the $y$ component of the momentum is a conserved quantity and can be expressed as $p_y = \hbar k_y$. The SOC contribution to $H_0$ in Eq.~(\ref{Ho}) can then be rewritten as
\begin{equation}
H_{SO}=\mathbf{w}(p_x,k_y)\cdot\boldsymbol{\sigma},
\end{equation}
where
\begin{equation}
\mathbf{w}(p_x,k_y)=\mathbf{w}(p_x)+\mathbf{w}(k_y)
\end{equation}
is the Rashba+Dresselhaus spin-orbit field, with
\begin{equation}\label{w-field-p}
\mathbf{w}(p_x)=\frac{p_x}{\hbar}
\begin{pmatrix}
\beta\cos 2\theta_c \\
-\alpha -\beta\sin2\theta_c \\
0
\end{pmatrix}
\end{equation}
and
\begin{equation}\label{w-field-k}
\mathbf{w}(k_y)=k_y
\begin{pmatrix}
\alpha -\beta\sin2\theta_c \\
-\beta \cos2\theta_c \\
0
\end{pmatrix}
\end{equation}
as the $p_x$- and $k_y$-dependent contributions, respectively.

Narrow junctions with Rashba SOC can be approximately described by an effective, one-dimensional BdG Hamiltonian that resembles the model Hamiltonian used for investigating MBS in semiconductor wires with proximity-induced superconductivity \cite{Lutchyn2010:PRL,Oreg2010:PRL}. Although such a simplified model cannot properly account for the crystalline anisotropy, it shows that the formation of robust MBS are favored when the external magnetic field is perpendicular to the Rashba spin-orbit field \cite{Lutchyn2010:PRL,Oreg2010:PRL}.

In the narrow junction limit, only the $k_y$-dependent contribution to the spin-orbit field is relevant and the spin-dependent interactions in the BdG Hamiltonian can be approximated as
\begin{equation}
H_{spin}\approx[\mathbf{w}(k_y)\cdot\sigma]\tau_z-\frac{g^\ast \mu_B}{2}\mathbf{B}\cdot\boldsymbol{\Sigma}.
\end{equation}
A rotation of the spin axes around the $z$-axis by an angle $\gamma$ transforms $\mathbf{w}(k_y)$ [see Eq.~(\ref{w-field-k})] back into a Rashba-like field $\tilde{\mathbf{w}}(k_y)$. Indeed,
\begin{equation}
U^\dagger [\mathbf{w}(k_y)\cdot\boldsymbol{\sigma}]\tau_z U=[\tilde{\mathbf{w}}(k_y)\cdot\boldsymbol{\sigma}]\tau_z,
\end{equation}
where $U=e^{-i\frac{\gamma}{2}\sigma_z}\otimes \sigma_0$,
\begin{equation}\label{u-def}
\tan\gamma =\frac{\beta\cos2\theta_c}{\beta\sin2\theta_c-\alpha},
\end{equation}
and $\tilde{\mathbf{w}}(k_y)=\tilde{\alpha}k_y(1,0,0)^T$
is the spin-orbit field in the spin rotated system. In the relations above $\sigma_0$ denotes the $(2\times 2)$ unit matrix and $\tilde{\alpha}=\sqrt{\alpha^2 +\beta^2 -2\alpha\beta\sin2\theta_c}$. Similarly, the spin rotation transforms the Zeeman interaction as,
\begin{equation}
U^\dagger(\mathbf{B}\cdot\boldsymbol{\Sigma})U=\tilde{\mathbf{B}}\cdot\boldsymbol{\Sigma},
\end{equation}
where
\begin{equation}\label{b-tilde}
\tilde{\mathbf{B}}=|\mathbf{B}|
\begin{pmatrix}
\cos(\theta_B - \theta_c - \gamma) \\
\sin(\theta_B - \theta_c - \gamma) \\
0
\end{pmatrix}.
\end{equation}

The unitary transformation $U$ leaves all the terms in the BdG Hamiltonian invariant, except the spin-dependent contribution,
\begin{equation}
U^\dagger H_{spin}U=[\tilde{\mathbf{w}}(k_y)\cdot\sigma]\tau_z-\frac{g^\ast \mu_B}{2}\tilde{\mathbf{B}}\cdot\boldsymbol{\Sigma}.
\end{equation}
Hence, $U$ transforms the BdG Hamiltonian with one-dimensional Rashba + Dresselhaus SOCs and a magnetic field with orientation angle $\theta_B$ into a similar problem but with an effective Rashba-like SOC, whose amplitude, $\tilde{\alpha}$, depends on the strength of the Rashba and Dresselahus fields, and the crystallographic orientation of the junction. In the spin-rotated system the orientation angle of the magnetic field is $\theta_B-\theta_c-\gamma$. Therefore, we expect that in the narrow junction limit, the optimal topological gap is realized when the magnetic field in the spin-rotated frame is perpendicular to the effective Rashba-like spin-orbit field, i.e., when $\tilde{\mathbf{w}}(k_y)\cdot\tilde{\mathbf{B}}=0$. This condition is realized when $\theta_B-\theta_c=\gamma+\pi/2$, or equivalently,
\begin{equation}\label{thb-thc}
\tan(\theta_B-\theta_c)=-1/\tan\gamma.
\end{equation}
Making use of Eqs.~(\ref{u-def}) and (\ref{thb-thc}) we obtain the relation,
\begin{equation}\label{bopt}
\theta_B =\theta_c + \arctan\left(\cot\theta_{so}\sec 2\theta_c-\tan 2\theta_c\right),
\end{equation}
where $\cot{\theta_{so}}=\alpha/\beta$.

Eq.~(\ref{bopt}) can be used to estimate the angle $\theta_B$ leading to the best topological protection, depending on $\theta_{so}$ and $\theta_c$. The optimal alignment of the magnetic field with respect to the junction crystallographic direction is crucial for the realization of robust MBS. As shown below (see also Fig.~\ref{fig:thc-phi}), although the TS state extents over a relatively large region of parameters, the existence of a sizable topological gap is reduced to small zones around the optimal alignment predicted by Eq.~(\ref{bopt}). Therefore, a small deviation of the magnetic field from its optimal orientation can lead to the collapse of the topological gap.  Experimental evidence of the high sensitivity of the TS state to the magnetic field orientation has recently been reported in Ref.~\cite{Mayer2021:PRL}.

For illustration, Fig.~\ref{fig:optimal} shows $\theta_B$, computed by Eq.~(\ref{bopt}), as a function of $\theta_c$ for different values of $\theta_{so}$. Without crystalline anisotropy ($\theta_{so}=n\pi/2$ with an integer $n$), i.e., if only Rashba or Dresselhaus SOC is present, the optimal orientation of $\mathbf{B}$ coincides with the BDI condition [Fig.~\ref{fig:optimal}(a)]. This implies that one cannot easily get free of the BDI class without quickly reducing $\Delta_{\rm top}$. In the presence of crystalline anisotropy, however, the optimal $\theta_B$ differs from the BDI condition [dashed lines in Fig.~\ref{fig:optimal}(b)], enabling a pure class D phase with finite $\Delta_{\rm top}$. In the special cases $\theta_{so}= (2n+1)\pi/4$ (i.e., $\alpha=\pm\beta$) the optimal $\theta_B$ is independent of $\theta_c$, therefore detuning $\theta_B$ from its optimal value results in a gapless spectrum independent of the crystallographic orientation [see Fig.~\ref{fig:thc-phi}(b)]. 

\begin{figure}[t]
\centering
\includegraphics*[width=8.5cm]{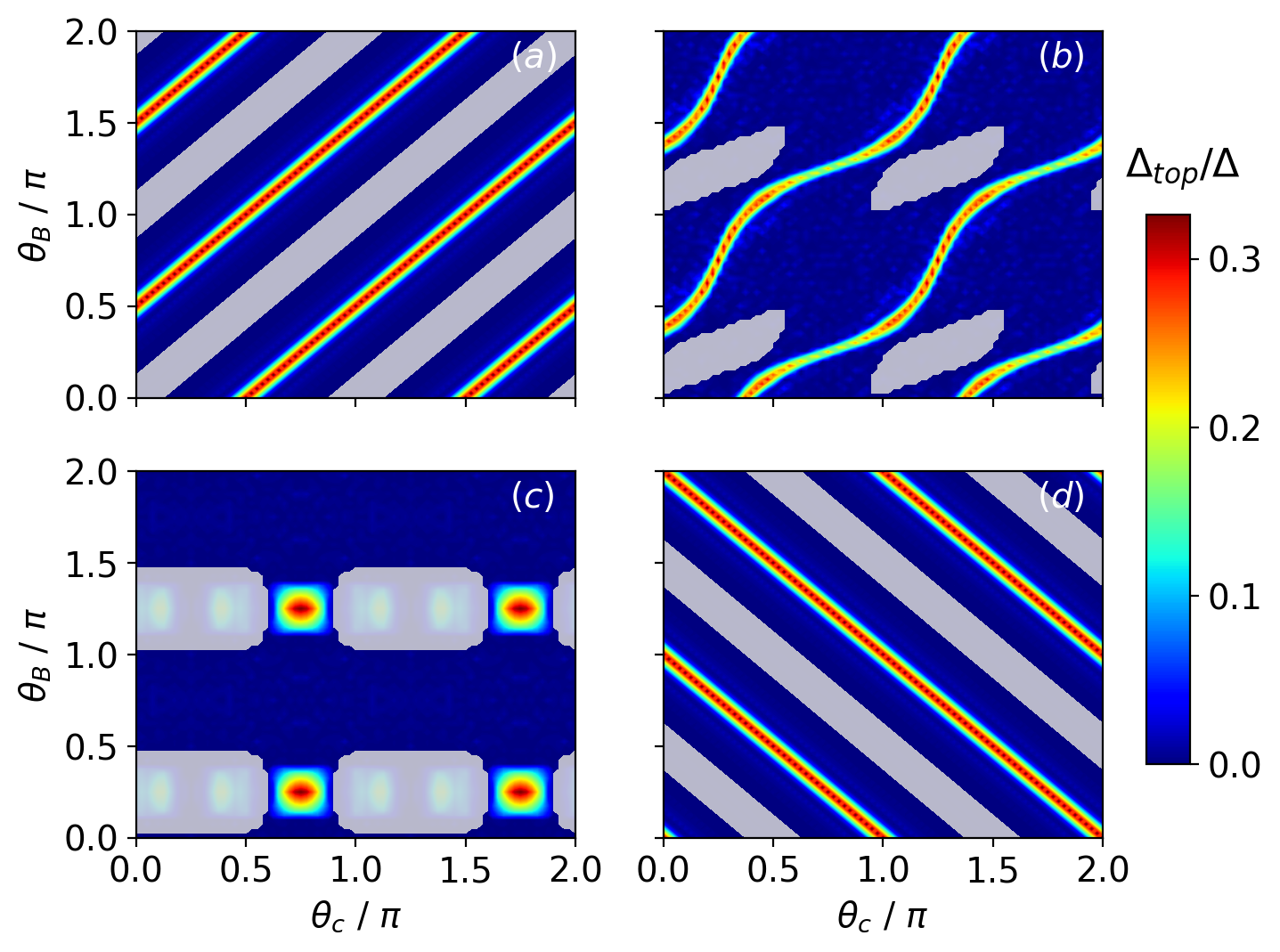}
\caption{Topological gap $\Delta_{\rm top}$ as a function of $\theta_c$ and $\theta_B$ for (a) $\theta_{so}=0$, (b) $\theta_{so}=\pi/8$, (c) $\theta_{so}=\pi/4$, and (d) $\theta_{so}=\pi/2$. $\phi=\pi$ in (a), (b), (d) and $\phi=0$ in (c). Gray-shaded areas correspond to trivial regions ($Q=1$).}\label{fig:thc-thb}
\end{figure}

This is corroborated by Fig.~\ref{fig:thc-thb}, which shows numerical calculations of $\Delta_{\rm top}$ that are in excellent agreement with the predictions of Eq.~(\ref{bopt}) shown in Fig.~\ref{fig:optimal}. Figure~\ref{fig:thc-thb} further shows that for a fixed orientation of $\mathbf{B}$ TS can still be controlled in junctions with different orientations by electrically tuning Rashba SOC (and thereby $\theta_{so}$). This can be relevant in more complex geometries, like zigzag-junctions~\cite{Laeven2020:PRL}, as well as in tree-junctions~\cite{Stenger2019:PRB,Stern2019:PRL,Yang2019:PRB}, and X-junctions~\cite{Zhou2019:PRL}, which have been proposed for fusing and braiding multiple Majorana pairs. Furthermore, since TS strongly depends on the crystalline anisotropy, small misalignments of $\theta_B$ and/or $\theta_c$ from the optimal configuration defined by Eq.~(\ref{bopt}) results in a fast decay of $\Delta_{\rm top}$. Therefore, the crystalline anisotropy can be used as an additional tool for experimentally distinguishing the topological nature of the gap-closing signatures observed in the critical current of planar JJs from non-topological effects such as Fraunhofer patterns and Fulde–Ferrell–Larkin–Ovchinnikov states, which exhibit a weaker dependence on $\theta_B$ and $\theta_c$.

\section{Topological Phase Transitions by Tuning the Spin-Orbit Coupling}

\begin{figure}[t]
	\centering
	\includegraphics*[width=8.5cm]{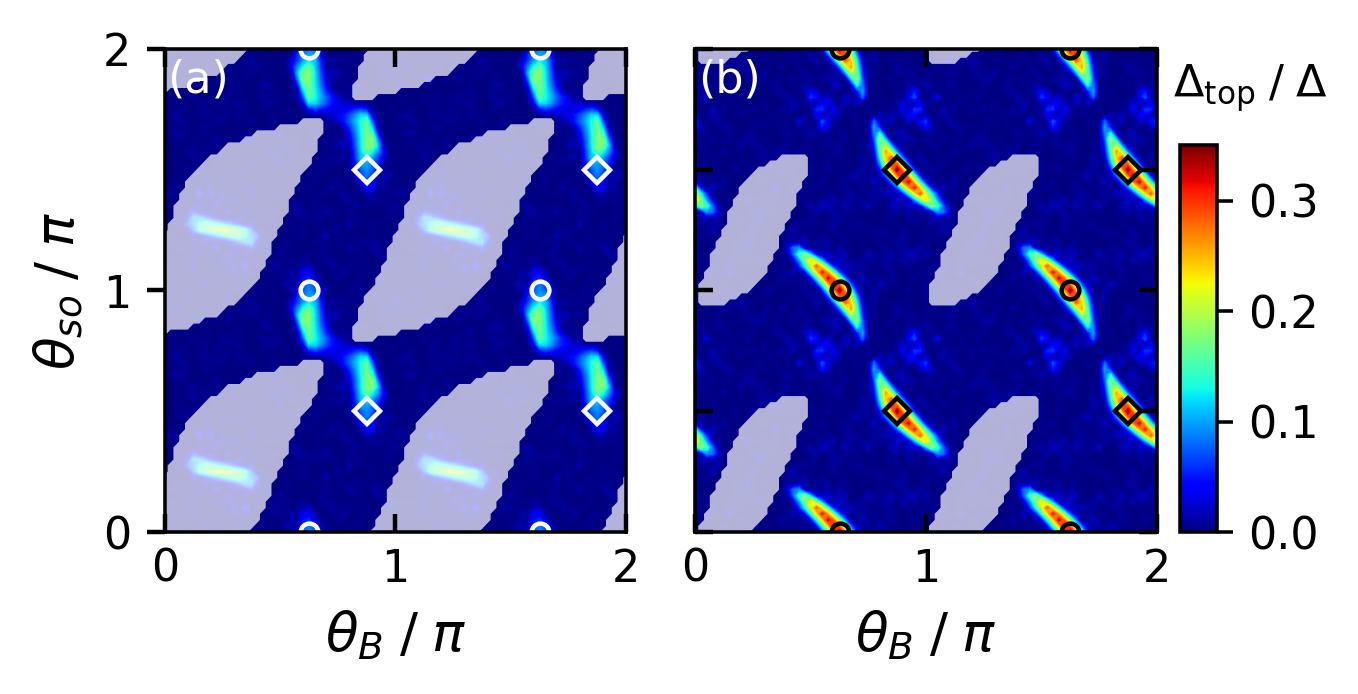}
	\caption{Topological gap $\Delta_{\rm top}$ as a function of $\theta_B$ and $\theta_{so}$ for $\theta_c=\pi/8$ and (a) $\phi=0$ and (b) $\phi=\pi$. Gray-shaded areas correspond to trivial regions ($Q=1$). The symbols indicate points at which the conditions for the BDI symmetry class (see Table~I in the main text) are fulfilled and correspond to cases in which only Rashba (circles) or only Dresselhaus (diamonds) are present.}\label{fig:thb-thso}
\end{figure}

A way to break the BDI symmetry consists in making the junction asymmetric with the two superconducting leads having different sizes and/or superconducting gap amplitudes. Although in such a case $\Delta_{\rm top}$ can remain finite, the tunability of the system is lost. However, controllable topological phase transitions between class D and class BDI TS can be achieved with CATS by electrically tuning the strength of $\alpha$ (and thus $\theta_{so}$). This can be inferred from Fig.~\ref{fig:thb-thso}, where $\Delta_{\rm top}$ is plotted as a function of $\theta_B$ and $\theta_{so}$ for fixed $\theta_c$ and different $\phi$. At $\phi=0$, the maximum topological gap resembles a more vertical pattern, and thus topological phase transitions between BDI and gapped D classes can be realized by just tuning $\theta_{so}$ once an optimal $\theta_{B}$ has been established [Fig.~\ref{fig:thb-thso}(a)]. However, at $\phi=\pi$ it is more convenient to tune both $\theta_{so}$ and $\theta_{B}$ in tandem, as shown in Fig.~\ref{fig:thb-thso}(b). The value of $\theta_{so}$ can be experimentally varied by tuning the strength of the Rashba SOC with the use of a top gate, as is customary in spintronic applications. The use of top gates for controlling the Rashba SOC strength in planar Josephson junctions has been experimentally demonstrated \cite{Mayer2021:PRL,Mayer2020:NC}.

\section{Conclusions}
Crystalline anisotropic topological superconductivity presents a promising path for manipulating Majorana bound states in phase-controlled planar Josephson junctions. The interplay between magnetic field, spin-orbit coupling and junction orientation allows for tuning and controlling the topological superconducting state (including transitions between BDI and D symmetry classes) and the localization of Majorana bound states. Our analytical formula for the optimal magnetic field orientation, confirmed by numerical simulations, can serve as a guide to future experiments seeking stable and well-localized Majorana bound states protected by a sizable topological gap.

\acknowledgments
\emph{Acknowledgments.} J.D.P. and A.M.A. acknowledge support from DARPA Grant No.
DP18AP900007 and US ONR Grant No. N000141712793.

\appendix
\section{}

To qualitatively understand the anisotropic effects on the localization length of the Majorana states we consider a junction with translational symmetry along the $y$ direction. In this case the wave function in the left superconducting region can be written as
\begin{widetext}
\begin{equation}\label{w-func}
\psi_L(x,y)=\frac{1}{\sqrt{L}}e^{ik_y}\sum_{s=\pm}\left[ A_{es}e^{iq_{es}(x+W_N/2)}\chi_{es}+ A_{hs}e^{-iq_{hs}(x+W_N/2)}\chi_{hs},\right],
\end{equation}
\end{widetext}
where the subindexes $e$ and $h$ refer to electron- and hole-like states, respectively, and $s$ characterizes the chirality of the Nambu spinors $\chi_{es}$ and $\chi_{hs}$. The Nambu spinors are the eigenvectors of the BdG Hamiltonian in Eq.~(\ref{H-BdG}) with
\begin{equation}
H_0=\xi+\mathbf{w}(k_y)\cdot\boldsymbol{\sigma}+\mathbf{w}(q)\cdot\boldsymbol{\sigma},
\end{equation}
where $\mathbf{w}(k_y)$ is given by Eq.~(\ref{w-field-k}),
\begin{equation}\label{wq}
\mathbf{w}(q)=q
\begin{pmatrix}
\beta\cos 2\theta_c \\
-\alpha-\beta \sin 2\theta_c \\
0
\end{pmatrix},
\end{equation}
and
\begin{equation}
\xi=\frac{\hbar^2(q^2+k_y^2)}{2m^\ast}-(\mu_S-\varepsilon).
\end{equation}

The BdG eigenproblem determines the eigenspinors $\chi$ and the dispersion relation expressing the eigenenergies as a function of the wave vector components $q$ and $k_y$. By requiring that $E\approx 0$ (as is the case for Majorana states) we obtain the following relation
\begin{widetext}
\begin{equation}\label{e-q}
(|\mathbf{J}|^2+|\mathbf{w}|^2+|\Delta|^2+\xi^2)^2
-4[|\mathbf{J}|^2(|\Delta|^2+\xi^2)+|\mathbf{w}|^2\xi^2+(\mathbf{w}\cdot\mathbf{J})^2]=0,
\end{equation}
\end{widetext}
where
\begin{equation}
\mathbf{w}=\mathbf{w}(q)+\mathbf{w}(k_y)\;\;\;;\;\;\;\mathbf{J}=\frac{g^\ast\mu_B}{2}\mathbf{B}.
\end{equation}

As can be deduced from Eq.~(\ref{w-func}), when the wave vectors $q_{e/h,s}$ (generically denoted by $q$) are purely real, the wave function represents propagating waves. However, when $q_{e/h,s}$ are complex, the wave function in the superconducting region decays within a length, $l$, inversely proportional to the imaginary part of $q$, i.e.,
\begin{equation}
l\sim 1/{\rm Im}[q].
\end{equation}
The values of $q$ and its imaginary part can be determined by calculating the roots of Eq.~(\ref{e-q}). The exact solutions can be found analytically, but the expressions are lengthy and not very illuminating. However, a qualitative understanding can be obtained by assuming $k_y\approx 0$ and neglecting terms of order higher than two in $q$. In such a case an approximate, simplified solution of Eq.~(\ref{e-q}) reads
\begin{widetext}
\begin{equation}\label{q-app}
qa\approx\frac{|\mathbf{J}|^2-(\mu_S-\varepsilon)^2-|\Delta|^2}{\sqrt{4(\mathbf{n}\cdot\mathbf{J})^2-4t(\mu_S-\varepsilon)[|\mathbf{J}|^2-(\mu_S-\varepsilon)^2-|\Delta|^2]-2|\mathbf{n}|^2[|\mathbf{J}|^2-(\mu_S-\varepsilon)^2+|\Delta|^2]}},
\end{equation}
\end{widetext}
where
\begin{equation}
\mathbf{n}=\frac{\mathbf{w}(q)}{qa}=\frac{1}{a}
\begin{pmatrix}
\beta\cos 2\theta_c \\
-\alpha-\beta \sin 2\theta_c \\
0
\end{pmatrix},
\end{equation}
$a$ is the lattice constant, and $t=\hbar^2/(2m^\ast a^2)$. The lattice constant has been introduced for convenience, so that $qa$ becomes dimensionless, but it has no actual influence in the value of $q$. It then follows from Eq.~(\ref{q-app}) that ${\rm Im}[q]\neq 0$ when
\begin{widetext}
\begin{equation}\label{c-e}
4(\mathbf{n}\cdot\mathbf{J})^2 < 4t(\mu_S-\varepsilon)[|\mathbf{J}|^2-(\mu_S-\varepsilon)^2-|\Delta|^2]+2|\mathbf{n}|^2[|\mathbf{J}|^2-(\mu_S-\varepsilon)^2+|\Delta|^2].
\end{equation}
\end{widetext}
In such a case the localization length is finite,
\begin{widetext}
\begin{equation}
\frac{l}{a}\approx\left|\frac{\sqrt{4t(\mu_S-\varepsilon)[|\mathbf{J}|^2-(\mu_S-\varepsilon)^2-|\Delta|^2]+2|\mathbf{n}|^2[|\mathbf{J}|^2-(\mu_S-\varepsilon)^2+|\Delta|^2]-4(\mathbf{n}\cdot\mathbf{J})^2}}{|\mathbf{J}|^2-(\mu_S-\varepsilon)^2-|\Delta|^2}\right|
\end{equation}
\end{widetext}
and can lead to the localization of the Majorana states at the ends of the junction. Conversely, if Eq.(\ref{c-e}) is not fulfilled, ${\rm Im}[q]=0$ and the localization length becomes infinite, i.e., the Majorana states spread along the edges perpendicular to the junction. Note that the fact that $l\rightarrow\infty$ is just an artifact of the assumption that the superconducting leads extend infinitely along the $x$-direction (otherwise $q$ would not be a good quantum number) but in reality the extension of the Majorana states is still limited by the sample size.

An important observation is that whether Eq.(\ref{c-e}) holds or not, depends on the value of $(\mathbf{n}\cdot\mathbf{J})^2$, which is a function of the junction and magnetic field directions, and the SOC strength. As a result, for some values of those parameters the localization length is small and the Majoranas are mainly localized at the ends of the junction, while for other values the localization length is enhanced and the Majoranas spread along the edges, as shown in Figs.~\ref{fig:spectrum}(b) and (d).

\section{}

\begin{figure}[t]
	\centering
	\includegraphics*[width=8.5cm]{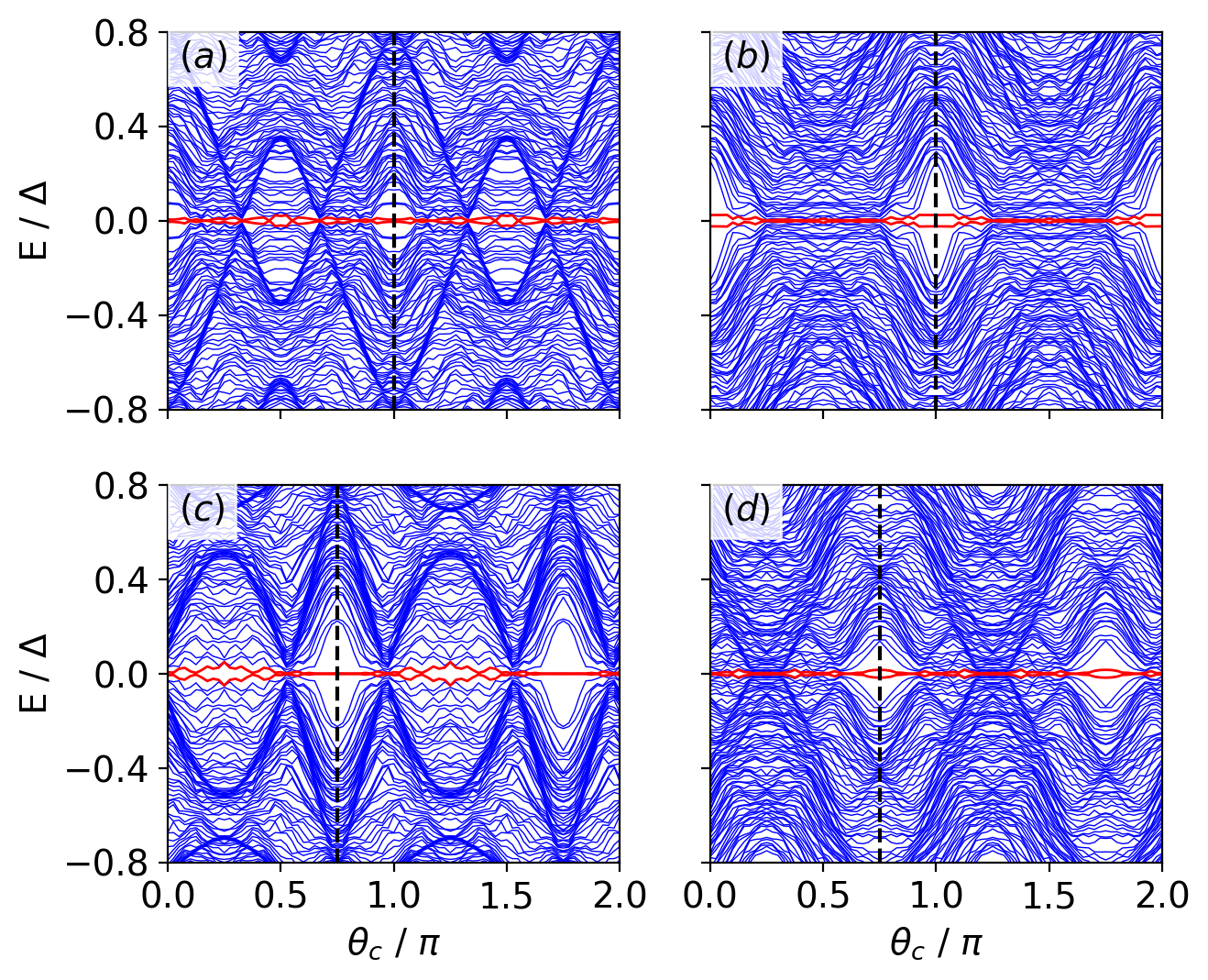}
	\caption{Energy spectrum as a function of the junction orientation, $\theta_c$ for (a) $\theta_{so}=\pi/2$ (only Dresselhaus SOC), $\theta_B=0$, and $\phi=0$, (b) $\theta_{so}=\pi/2$ (only Dresselhaus SOC), $\theta_B=0$, and $\phi=\pi$, (c) $\theta_{so}=\pi/8$, $\theta_B=\pi/4$, and $\phi=0$, and (d) $\theta_{so}=\pi/8$, $\theta_B=\pi/4$, and $\phi=\pi$. The red lines indicate the two states with energies closest to zero.}\label{fig:e-thc}
\end{figure}

Complementary to Fig.~\ref{fig:thc-phi}(c) in the main text, we show in Figs.~\ref{fig:e-thc}(a) and (b) the energy spectrum as a function of the junction crystallographic orientation, $\theta_c$ for a magnetic field orientation, $\theta_B=0$, i.e., along the $[100]$ direction. Figures \ref{fig:e-thc}(a) and (b) correspond to a junction in which only Dresselhaus SOC is present (i.e., $\theta_{so} =\pi/2$) and the phases are $\phi=0$ and $\phi =\pi$, respectively. The red lines indicate the two states with energy closest to zero. Although the energy of these states remain close to zero for all the junction orientations, the topological gap protecting the states is appreciable only in small regions in the vicinity of certain junction directions and is larger for $\phi=\pi$ (b). A similar behavior is observed when both Rashba and Dresselhaus SOCs are present. This is illustrated in Figs.~\ref{fig:e-thc}(c) and (d), which serves as a complement to Fig.~\ref{fig:thc-phi}(e) in the main text. In this case, however, the topological gap is larger for $\phi=0$ (c).

Figure \ref{fig:e-thc}, when compared to Fig.~\ref{fig:thc-phi} in the main text, demonstrates that although the parameter domain containing the topological state may be relatively large, the topological gap protecting the Majorana states is sizable only in a reduced parameter subspace. Therefore, an appropriate analysis of the effects of crystalline anisotropy is crucial for designing optimal experimental setups for realizing and detecting topological superconductivity in planar JJs.

\begin{figure}[t]
	\centering
	\includegraphics*[width=8.5cm]{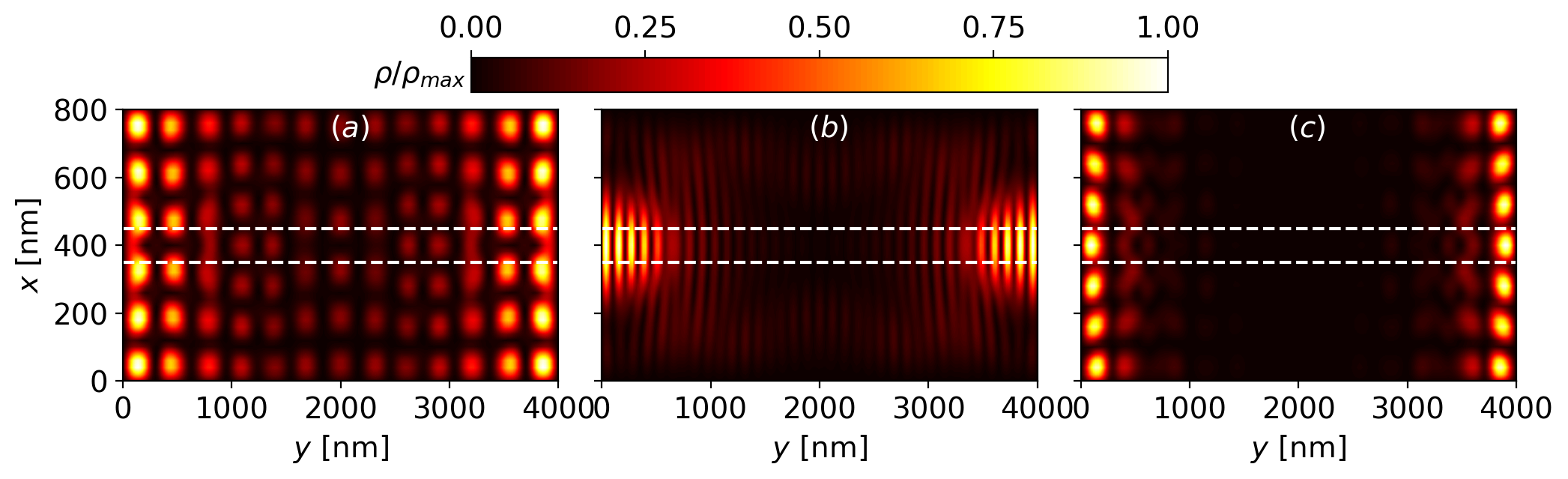}
	\caption{Probability density (normalized to its maximum value) of states with energies closest to zero (see red lines in Fig.~\ref{fig:e-thc}). Panels (a)-(c) correspond to the red-line states at the junction orientations indicated by black dashed lines in Figs.~\ref{fig:e-thc}(a)-(c), respectively. (a) $\theta_{so}=\pi/2$ (only Dresselhaus SOC), $\theta_B=0$, $\phi=0$, and $\theta_c=\pi$, (b) $\theta_{so}=\pi/2$ (only Dresselhaus SOC), $\theta_B=0$, $\phi=\pi$, and $\theta_c=\pi$, and (c) $\theta_{so}=\pi/8$, $\theta_B=\pi/4$, $\phi=0$, and $\theta_c=3\pi/4$. The white dashed lines indicate the edges of the junction channel. When $\phi=0$ the Majorana states are localized along the edges perpendicular to the junction channel, as shown in (c). However, for $\phi=\pi$ the Majorana states are mainly localized at the end regions inside the junction channel [see (b)].}\label{fig:p-density}
\end{figure}

The probability density (normalized to its maximum value) is shown in Figs.~\ref{fig:p-density}(a)-(c) for the states with energies closest to zero in correspondence with Figs.~\ref{fig:e-thc}(a)-(c), respectively. In each case, the junction orientation was set to the values indicated by black dashed lines in Fig.~\ref{fig:e-thc}. When only Dresselhaus SOC is present (a similar behavior occurs when only Rashba SOC is present), the formation of robust Majorana states is favored when $\phi=\pi$. Indeed, as shown in Fig.~\ref{fig:p-density}(a) the states with energies closest to zero exhibit a poor localization when $\phi=0$. This is a consequence of an extended wave function overlapping and a small topological gap. However, well localized Majorana states form when $\phi=\pi$ [see Fig.~\ref{fig:p-density}(b)].

In the presence of crystalline anisotropy (i.e., when both Rashba and Dresselhaus SOCs are present), it is possible to create well localized Majorana states with a sizable topological gap when $\phi=\pi$ but also when $\phi=0$ [see Fig.~\ref{fig:p-density}(c)]. However, the localization of the Majorana states appears to be different in dependence on whether the phase difference is $0$ or $\pi$. When $\phi=0$ the Majorana states are localized along the edges perpendicular to the junction channel [see Figs.~\ref{fig:p-density}(c)], while for $\phi=\pi$ the states are mainly localized at the end regions, inside the junction channel [see Fig.~\ref{fig:p-density}(b)].


%

\end{document}